\documentclass[preprint,review,12pt]{elsarticle}

\usepackage{amssymb}

\usepackage{hyperref}
\usepackage{amsmath}
\usepackage{cleveref}
\usepackage{booktabs}
\usepackage{caption,setspace}
\usepackage{tabularx}
\usepackage[export]{adjustbox}
\usepackage{tikz}
\usepackage{geometry}
\usepackage{wrapfig} 
\usepackage{flushend} 
\usepackage{algorithm}
\usepackage{algpseudocode}

\usepackage{helvet}  

\def\checkmark{\tikz\fill[scale=0.3](0,.45) -- (.25,0) -- (.8,.62) -- (.25,.15) -- cycle;} 
\hypersetup{pdfauthor={Name}}

\makeatletter
\newcommand{\ssymbol}[1]{^{\@fnsymbol{#1}}}
\makeatother

\usepackage{array}
\newcolumntype{P}[1]{>{\centering\arraybackslash}p{#1}}

\AtBeginDocument{
    \crefformat{equation}{#2Eq.~#1#3}
    \crefformat{figure}{#2Fig.~#1#3}
    \crefformat{table}{#2Tab.~#1#3}
}

\journal{arXiv} 

\begin{document}
\begin{sloppypar}

\begin{frontmatter}

\title{GSDA: Generative Adversarial Network-based Semi-Supervised Data Augmentation for Ultrasound Image Classification}


\author[a]{Zhaoshan Liu}
\ead{e0575844@u.nus.edu} 

\author[a,b]{Qiujie Lv}
\ead{lvqj5@mail2.sysu.edu.cn} 

\author[c]{Chau Hung Lee}
\ead{chau_hung_lee@ttsh.com.sg}

\author[a]{Lei Shen\corref{cor1}}
\ead{mpeshel@nus.edu.sg} 
\cortext[cor1]{Corresponding author}

\address[a]{Department of Mechanical Engineering, National University of Singapore, 9 Engineering Drive 1, Singapore, 117575, Singapore}

\address[b]{School of Intelligent Systems Engineering, Sun Yat-sen University, No.66, Gongchang Road, Guangming District, 518107, China}

\address[c]{Department of Radiology, Tan Tock Seng Hospital, 11 Jalan Tan Tock Seng, Singapore, 308433, Singapore}


\begin{abstract}
Medical Ultrasound (US) is one of the most widely used imaging modalities in clinical practice, but its usage presents unique challenges such as variable imaging quality. Deep Learning (DL) models can serve as advanced medical US image analysis tools, but their performance is greatly limited by the scarcity of large datasets. To solve the common data shortage, we develop GSDA, a Generative Adversarial Network (GAN)-based semi-supervised data augmentation method. GSDA consists of the GAN and Convolutional Neural Network (CNN). The GAN synthesizes and pseudo-labels high-resolution, high-quality US images, and both real and synthesized images are then leveraged to train the CNN. To address the training challenges of both GAN and CNN with limited data, we employ transfer learning techniques during their training. We also introduce a novel evaluation standard that balances classification accuracy with computational time. We evaluate our method on the BUSI dataset and GSDA outperforms existing state-of-the-art methods. With the high-resolution and high-quality images synthesized, GSDA achieves a 97.9\% accuracy using merely 780 images. Given these promising results, we believe that GSDA holds potential as an auxiliary tool for medical US analysis.
\end{abstract}



\end{frontmatter}



\section{Introduction}\label{1}

Medical Ultrasound (US) has become a widely utilized screening and diagnostic tool in clinical practice due to its absence of ionizing radiation, high sensitivity, portability, and relatively low cost \citep{zlitni2018molecular}. However, there are limitations to be solved. Image quality is easily affected by noise and artifacts, inter-operator variability is considerable, and variability across different US systems is usually high. Due to these, diagnosing medical US images always heavily relies on radiologists. To address the problem, developing an advanced medical US image analysis tool to make medical US diagnosis more objective, accurate, and automatic is essential. In recent years, Deep Learning (DL) has emerged as a powerful tool to automate the extraction of useful information from big data. It has enabled ground-breaking advances in numerous computer vision tasks \citep{voulodimos2018athanasios}. For the classification task, the Convolutional Neural Network (CNN) \citep{lecun1998gradient} is one of the most dominant methods. However, effectively training a CNN typically requires large datasets, which are often a significant obstacle in the medical field. For one thing, the acquisition of medical images typically necessitates the use of specialized equipment and requires medical experts for annotation. For another, datasets are usually confidential due to privacy concerns. In this case, the Transfer Learning (TL) technique is widely implemented in CNN to relieve the common data shortage. The TL allows the models to be trained on the larger dataset first to relieve the difficulty of model training. However, given the unique characteristics of US images and the acute data shortage, relying solely on TL often fails to guarantee the model performance \citep{al2019deep}.

To further improve the model performance for US image classification, various Data Augmentation (DA) methods have been widely adopted. Traditional DA methods typically generate images based on a sequence of transformations such as rotation, flip, etc. While such methods are beneficial, they come with several challenges. Firstly, manually designing the type and sequence of transformations largely depends on experience and can often lead to suboptimal results. Secondly, the number of combinations is restricted when leveraging a small number of transformations. Although expanding the number of transformations can potentially address this, excessive transformations might produce meaningless augmented images that drift significantly from the original \citep{hendrycks2019augmix}. Due to this, several advanced DA methods are proposed \citep{yang2022image, liu2023medaugment} and the Generative Adversarial Network (GAN) \citep{goodfellow2014generative} is one of the most widely implemented ones. The GAN is widely implemented for medical image synthesis and is composed of a generator ($G$) and a discriminator ($D$) playing an adversarial “game”. During training, the $G$ synthesizes images based on the data distribution it learned, and the $D$ tries to discriminate whether the images are synthesized or not.

Several previous works \citep{madani2018semi, pang2021semi, al2019deep} have utilized GAN for DA and classified medical images in a semi-supervised way. However, there are several points to be improved. For one thing, the synthesized images have either low resolution or quality. This can be caused by the basic GAN structure as well as the lack of the TL technique during training GAN. For another, the quality of the synthesized images is not evaluated quantitatively and the data distribution relationship across real and synthesized images is not investigated. Besides, performance across different CNN models is not fully searched. Till now, synthesizing high-resolution and high-quality US images, as well as training a high-performance classification model with a small dataset remain challenging. To solve existing problems, here we propose GSDA consisting of CNN and GAN. The GAN synthesizes and pseudo-labels the artificial US images with high resolution and high quality, whereas the CNN is trained using both real and synthesized images. To enhance image resolution and quality, we adopt the state-of-the-art GAN model SGA \citep{karras2020training} and employ the TL technique during its training. To evaluate the synthesized images quantitatively, we implement widely accepted standards Inception Score (IS) \citep{salimans2016improved} and Fréchet Inception Distance (FID) \citep{heusel2017gans}. We also implement t-SNE for analyzing the data distribution across real and synthesized images. To fully search the performance across different CNN models, we implement intensive experiments on several CNN models and compare the results. Moreover, we also propose a novel evaluation standard, the Training Efficiency Index (TEI), to balance the accuracy and the training time consumption. We evaluate our GSDA on the BUSI dataset \citep{al2020dataset}, and the results show that with high-resolution and high-quality images synthesized, our GSDA can obtain a 97.9\% accuracy using merely 780 images. To sum up, our main contributions are:
\begin{itemize}
    \item We propose a GAN-based semi-supervised DA method GSDA to solve the common data shortage.
    \item We leverage state-of-the-art GAN to synthesize high-resolution and high-quality US images.
    \item We evaluate the synthesized image quantitatively and analyze the data distribution between real and synthesized images.
    \item We propose a novel evaluation standard to balance the classification accuracy and the time consumption.
\end{itemize}

The rest of this paper is organized as follows: In Section \hyperref[2]{2}, we illustrate the related works of GAN as well as its application on semi-supervised medical image classification. The description of the datasets used, together with the methods proposed are discussed in Section \hyperref[3]{3}. Section \hyperref[4]{4} shows the core experiment results, detailed analysis, and extensive ablation study. We conclude our work and point out the future perspective in Section \hyperref[5]{5}.


\section{Related work}
\label{2}

\textbf{GAN}. Many variants of GAN \citep{isola2017image, zhu2017unpaired, kim2019u, mirza2014conditional, radford2016unsupervised, gulrajani2017improving, karras2020analyzing} have been developed since it was initially proposed. In 2016, Radford et al. \citep{radford2016unsupervised} developed a DCGAN model, in which the convolution operation is introduced into the GAN. In DCGAN, both the $D$ and $G$ are trained once during each epoch. One year later, the WGAN was proposed by Arjovsky et al. \citep{arjovsky2017wasserstein}, which employs the Wasserstein distance into GAN and uses RMSprop as the optimizer instead of Adam. A variant of it, WGAN-GP, was later proposed \citep{gulrajani2017improving}. The WGAN-GP adds a gradient penalty and applies layer norm \citep{ba2016layer} in $D$. However, training these GAN models always needs a large number of images. Besides, the resolution of synthesized images is relatively low. In 2020, Karras et al. \citep{karras2020training} developed a novel SGA network with advanced architectural design, in which Adaptive Data Augmentation (ADA) is introduced in GAN to handle small data regimes. To effectively handle high-resolution images, both the $G$ and $D$ of the SGA are designed with a hierarchical structure.

\textbf{GAN-based semi-supervised medical image classification}. To overcome the common data shortage in the field of medical image classification, several works \citep{madani2018semi, amin2020semi, pang2021semi, al2019deep, frid2018gan} have been proposed to use GAN for DA and classify the images in a semi-supervised way. The existing works can be divided into two approaches. The first is to train GAN solely and use the $D$ of GAN as a classifier \citep{madani2018semi, amin2020semi}. The second is to train GAN first and then use separate CNN as a separate classifier \citep{pang2021semi, al2019deep, frid2018gan}. We opt for the latter approach, as it allows us to employ multiple CNN models and compare their performances. Compared with existing methods, our GSDA has several advantages. First, instead of employing basic GAN models, we implement state-of-the-art GAN model SGA to synthesize images with higher resolution and quality. We observed that most of the existing work using GAN does not introduce the TL technique thus hampering the model performance. We thus implement the TL technique rather than solely training from scratch. Second, besides qualitatively observing the synthesized quality, we employ IS and FID to evaluate the synthesized images quantitatively. We also visualize and analyze the data distribution across real and synthesized images. Third, we implement intensive experiments across different CNN models to search for higher performance. Finally, we propose a new evaluation standard TEI to balance the classification accuracy and the time consumption.


\section{Materials and methods}
\label{3}

\subsection{Datasets}
\label{3.1}

\begin{figure*}
	\centering
		\includegraphics[width=0.84\textwidth]{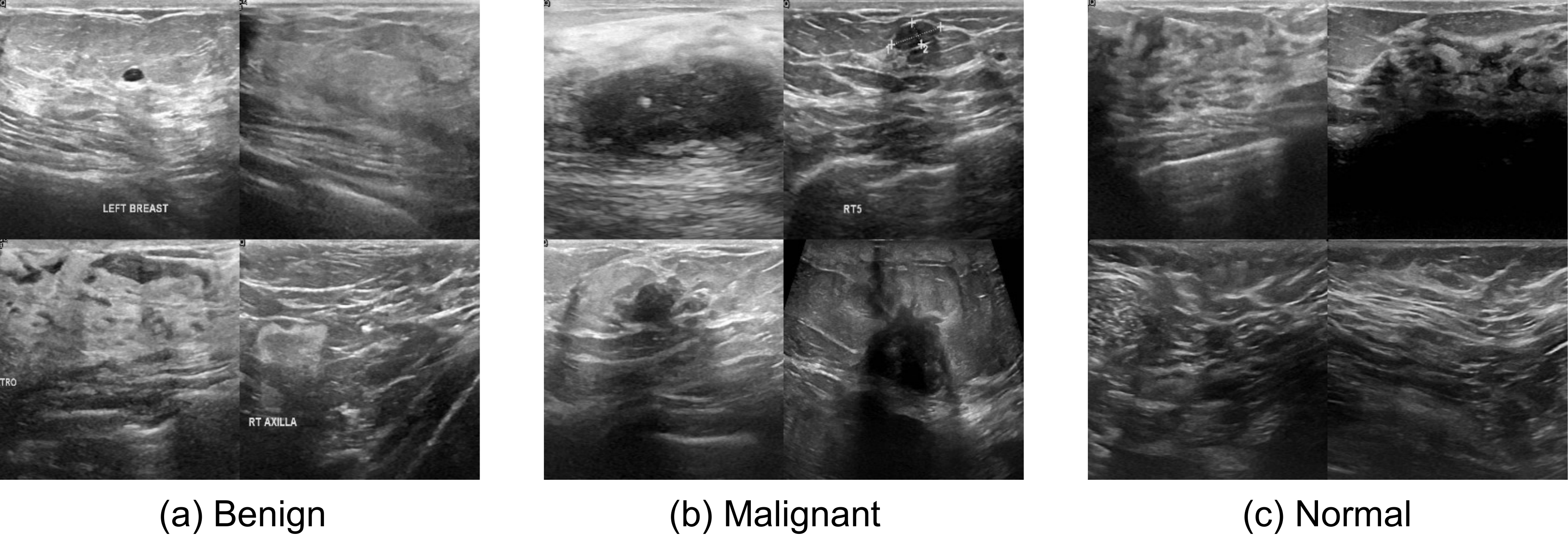}
	  \caption{Example figures accessed from the BUSI dataset. (a) benign, (b) malignant, and (c) normal.}
   \label{fig1}
\end{figure*}

We use the BUSI dataset for training and several example figures accessed from it are illustrated in \Cref{fig1}. The BUSI dataset is a breast cancer dataset collected among 600 female patients between 25 and 75 years old in 2018. The data is collected using the LOGIQ E9 US and LOGIQ E9 Agile US systems. The BUSI dataset contains 780 images and is divided into three subsets, including benign, malignant, and normal, as illustrated in \Cref{fig1}a, \Cref{fig1}b, and \Cref{fig1}c, respectively. Each subset corresponds to different breast cancer conditions. The benign, malignant, and normal subsets contain 437, 210, and 133 images, respectively. The average resolution of the images is around 500 $\times$ 500. Besides the BUSI dataset, four large datasets are selected as the source datasets of the TL technique. For synthesis, Flickr-Faces-HQ (FFHQ) \citep{karras2019style}, Large-Scale CelebFaces Attributes (CelebA) \citep{karras2017progressive}, and Large-scale Scene Understanding Challenge (LSUN) DOG \citep{yu2015lsun} are leveraged. FFHQ is a face dataset with 70K images at the resolution of 1024 × 1024. There is considerable variation in age, ethnicity, and image background among all images. CelebA is a large-scale face attributes dataset with 200K celebrity face images collected from 10,177 identities. Large pose variations and background clutter are covered. LSUN DOG contains 5M images of the category dog. For classification, ImageNet \citep{deng2009imagenet} is utilized. ImageNet, with its 14M images, is organized according to the nouns of the WordNet hierarchy, with each node represented by numerous images.

\subsection{SGA}
\label{3.2}

We leverage SGA to synthesize medical US images. The SGA features a $G-D$ architecture and includes an ADA block. Both $G$ and $D$ in the SGA follow a hierarchical structure, in which the resolution for $G$ progresses from low to high and vice versa for $D$. The detailed structure of $G$ and $D$ of SGA can be found in \Cref{fig2}b. The ADA is composed of eighteen transformations grouped into six groups, including pixel blitting, more general geometric transformations, color transforms, image-space filtering, additive noise \citep{sonderby2016amortised}, and cutout \citep{devries2017improved}. The set of transformations is employed in a fixed order with a strength $p \in[0,1]$. The $p$ is adaptively controlled based on the degree of overfitting. The evaluation of overfitting is to utilize a separate validation set and observe its behavior with respect to the training set. Let us denote the outputs of $D$ by $D_{\text {train }}, D_{\text {validation }}$, and $D_{\text {synthesized }}$, for the training set, validation set, and synthesized images, respectively, and their mean over $N$ consecutive batches by $\mathbb{E}[\cdot]$, the overfitting can be computed using the below equation:
\begin{equation}
r_v=\frac{\mathbb{E}\left[D_{\text {train }}\right]-\mathbb{E}\left[D_{\text {validation }}\right]}{\mathbb{E}\left[D_{\text {train }}\right]-\mathbb{E}\left[D_{\text {synthesized }}\right]} \quad r_t=\mathbb{E}\left[\operatorname{sign}\left(D_{\text {train }}\right)\right]
\label{eq1}
\end{equation}
where $r=0$ represents no overfitting and $r=1$ shows completely overfitting. $r_v$ shows the output for the validation set relative to the training set and synthesized images, and $r_t$ estimates the portion of the training set with positive $D$ outputs. For the adaptively controlled $p$, it is initialized to zero and adjusted once every four mini batches based on the \Cref{eq1}. In case the results indicate too much/little overfitting occurs, the $p$ is adjusted by incrementing/decrementing a fixed amount. 

The three subsets of the BUSI dataset are each used to train the SGA. Real images are preprocessed to a resolution of 256 $ \times $ 256. The resolution of synthesized images is also set as 256 $ \times $ 256 to balance the image quality and time consumption.  The loss function implemented is the non-saturating logistic loss $f(x)=\log (\operatorname{sigmoid}(x))$ \citep{goodfellow2014generative}. The $D$ loss is computed as $-f(x)$, where the $G$ loss is computed using $-f(x)$ and $-f(-x)$. The optimizer is Adam and the learning rate is 0.0025. The number of iterations is 4000 with a batch size of 32. SGA is trained using four different settings, including training from scratch and using the TL technique with three different source datasets to demonstrate the impact of the TL technique.

\subsection{GSDA}
\label{3.3}

\begin{figure*}
	\centering
		\includegraphics[width=\textwidth]{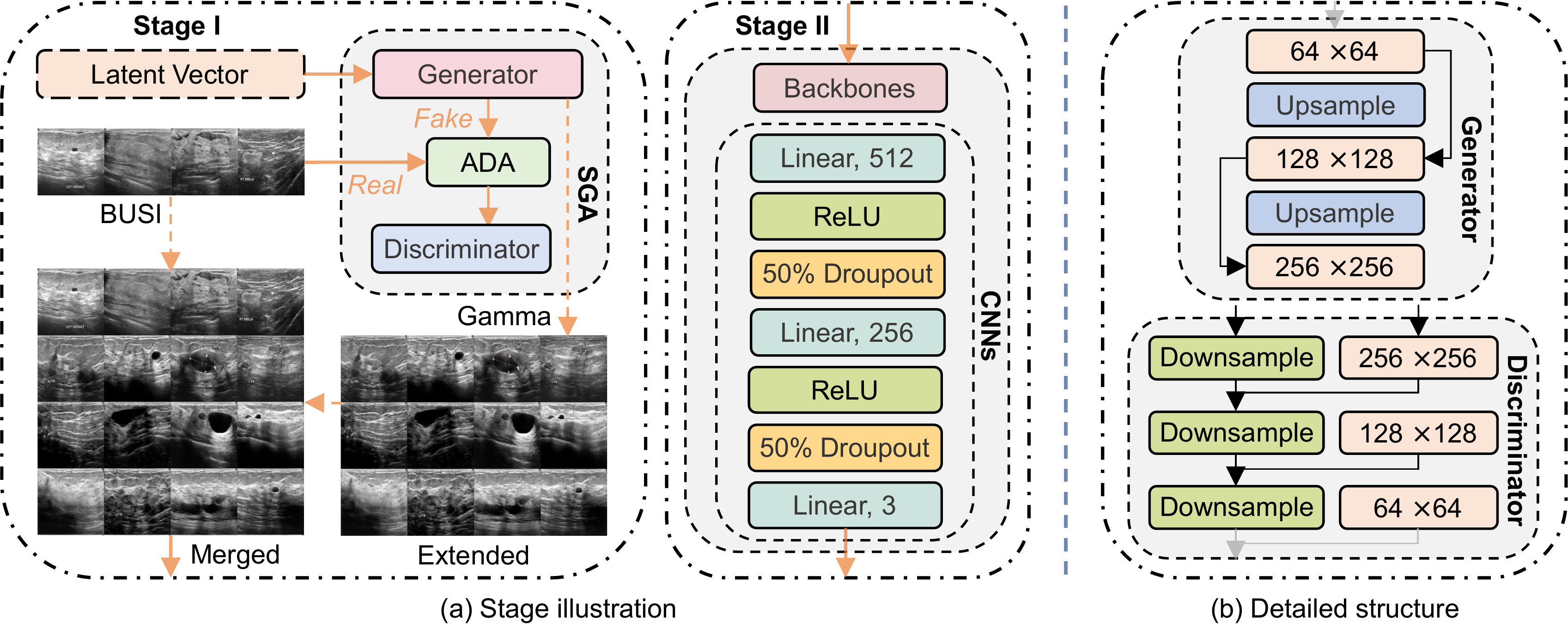}
	  \caption{The proposed GSDA is composed of two stages. In the first stage, SGA is trained using the BUSI dataset to capture the real image data distribution and synthesize artificial images. The synthesized extended datasets are then merged with the BUSI dataset to compose merged datasets. In the second stage, different CNN models are trained using the merged datasets. ADA stands for adaptive discriminator augmentation. Solid and dotted orange arrows show the training stream and data stream, respectively. Grey arrows point toward the omitted similar structures at different resolutions. (a) stage illustration, and (b) detailed structure of $G$ and $D$ of SGA.}
   \label{fig2}
\end{figure*}

As shown in \Cref{fig2}a, our GSDA is composed of two stages. In the first stage, we train SGA using the BUSI dataset to capture the real image data distribution for image synthesis. In the second stage, we train different CNN models using the merged datasets. We construct seven CNN models with the unique classification head using VGGNet \citep{simonyan2015very}, ShuffleNet \citep{zhang2018shufflenet}, ResNeXt \citep{xie2017aggregated}, ResNet \citep{he2016deep}, MobileNet \citep{howard2019searching}, InceptionNet \citep{szegedy2016rethinking}, and DenseNet \citep{huang2017densely} as the backbone. The classification head consists of two linear layers with a ReLU activation function, two dropout layers, and a linear layer with three output nodes, which equals the number of subsets. It is worth noting that we denote the combination of the SGA and different CNN models used in different groups of experiments as different SGA-CNN pairs. This results in seven SGA-CNN pairs, which are SGA-VGG, SGA-Shuffle, SGA-ResNeXt, SGA-Res, SGA-Mobile, SGA-Inception, and SGA-Dense.

We utilize SGA to synthesize medical US images. We endow the synthesized images with pseudo-labels that are the same as the real images. Specifically, the images synthesized by the SGA trained with the benign subset are pseudo-labeled benign, and the same process is performed for the malignant subset and the normal subset. We use the synthesized images to compose the extended datasets. The size of each extended dataset equals the integer $ \gamma $ multiple of the BUSI dataset. To keep class balance, the proportion of the three subsets in the extended datasets is the same as that of the BUSI. For each SGA-CNN pair, the maximum value of $ \gamma $ is determined experimentally based on our proposed evaluation standard TEI. For the detailed algorithm on how the maximum value of $ \gamma $ is determined, see Section \hyperref[3.4]{3.4}. We add the extended datasets to the BUSI dataset to compose the merged datasets. 

We train pre-trained CNN models using the merged datasets. The merged datasets are divided randomly into a training set and a validation set with a ratio of 8:2. The resolution of images is preprocessed to 224 $ \times $ 224 or 299 $ \times $ 299 (InceptionNet) due to different model architecture designs. The loss function is cross-entropy. The optimizer used is Adam and the learning rate equals 0.003. The weight decay (WD) is set to 0.001 if applicable. The number of epochs is 60 and the batch size is 32. For a given $ \gamma $, each CNN is trained under two groups of settings, depending on whether the WD is chosen or not. Traditional DA methods, including RandomResizedCrop and RandomHorizontalFlip, are implemented.

\subsection{Evaluation standards}
\label{3.4}

To evaluate the quality of images synthesized by the SGA, we use two ways of evaluation. The first is through qualitative observation where the overall quality and basic details are directly distinguishable. The second is quantitative assessment using IS and FID. The IS and FID are two prevalent evaluation standards for image synthesis and can be calculated via:
\begin{equation}
IS=\exp \left(\mathbb{E}_{x \sim p_{g}} D_{K L}(p(y \mid x) \| p(y))\right)
\label{eq2}
\end{equation}
\begin{equation}
FID=\left\|{m}-{m}_{w}\right\|_{2}^{2}+Tr\left({C}+{C}_{w}- 2\left({C} {C}_{w}\right)^{1 / 2}\right)
\label{eq3}
\end{equation}
where $ x \sim p_{g} $ represents sample $ x $ from $ p_{g} $, $ D_{K L} $ represents the KL divergence. $ w $ represents the real-world data, $ m $ denotes the mean value, $ C $ shows the covariance matrix, and $ Tr $ represents the trace. A lower IS indicates worse model performance, whereas a lower FID indicates better performance. From \Cref{eq2} and \Cref{eq3}, it is observed that the real images are not taken into consideration when calculating IS. In such scenarios, the model might achieve a high IS by simply replicating the real images. To make the evaluation more convincing, we regard FID as the main standard and take IS as a reference. Both FID and IS are calculated every 200 iterations. 

For the SGA-CNN pair, the training time $ t_{\gamma} $  increases significantly with $ \gamma $. We thus evaluate the model performance in two-fold. First, we employ classification accuracy $ acc_{\gamma} $ for evaluation. This is the most distinct and effective evaluation standard. Second, we propose a new standard TEI to balance the classification accuracy and the time consumption. Given $ \gamma $, TEI can be calculated using:
\begin{equation}
TEI= \begin{cases}\ln \left(t_{\gamma}-t_{\gamma=0}\right)^{-1} \cdot \left(acc_{\gamma}-acc_{\gamma=0}\right) & \gamma > 1, \\ 0 & \gamma = 0 .\end{cases}
\label{eq4}
\end{equation}
where $ t_{\gamma=0} $, $ acc_{\gamma=0} $ denotes the training time and accuracy when training using the BUSI dataset. From \Cref{eq4}, we can find that the TEI indicates the ability of the model to attain improved accuracy within a limited training time. The higher the TEI, the better the ability. It is worth noting that for each SGA-CNN pair, we determine the maximum value of $ \gamma $ experimentally based on the proposed TEI. The specific procedure is (1) Initialize $ \gamma $ as 1, (2) calculate TEI, (3) increase $ \gamma $ by 1, (4) calculate new TEI, (5) compare new TEI with the previous one, (6) if TEI increases, repeat (3), (4), and (5) until TEI stops increasing. The pseudo-code of the proposed algorithm is illustrated in \Cref{algorithm1}.

\begin{algorithm}[!pt] 
\begin{algorithmic}[1]
\caption{Determination of the maximum value of $ \gamma $ for SGA-CNN pairs.}
\label{algorithm1}  
\Require Extended multiple $\gamma \in[0,n]$, GAN model $ SGA $, $ i_{th} $ CNN model $ CNN_i, i \in[1,7]$, dataset $D=(X_{image}, Y)$, time $t$;
\Ensure $ TEI_{max}\mid\gamma $; 
\State $ TEI_{\gamma=0} = 0 $;
\ForAll {$i$}
    \State Initialize $ \gamma = 1 $;
    \While {$TEI_{\gamma,i} > TEI_{\gamma-1,i} $}
        \State $ t_{start} = start $ $ time $;
        \State $ X^{SGA}_{image} = SGA(X_{image}) $;
        \State $ y = CNN_i(X^{SGA}_{image} \bigcup X_{image}) $;
        \State $ acc_{\gamma, i} = accuracy(y, Y) $;
        \State $ t_{end} = end $ $ time $;
        \State $ t_{\gamma, i} = t_{end} - t_{start} $;
        \State Compute $TEI_{\gamma, i}=\ln \left(t_{\gamma,i}-t_{\gamma-1,i}\right)^{-1} \cdot \left(acc_{\gamma,i}-acc_{\gamma-1,i}\right)$;
        \State $ \gamma = \gamma + 1 $;
    \EndWhile
    \State Out $ \gamma $ when $TEI_{max}\mid\gamma$;
\EndFor
\end{algorithmic}
\end{algorithm}


\section{Results and analysis}
\label{4}

\subsection{Unsupervised synthesis}
\label{4.1}

\begin{figure*}
	\centering
		\includegraphics[width=\textwidth]{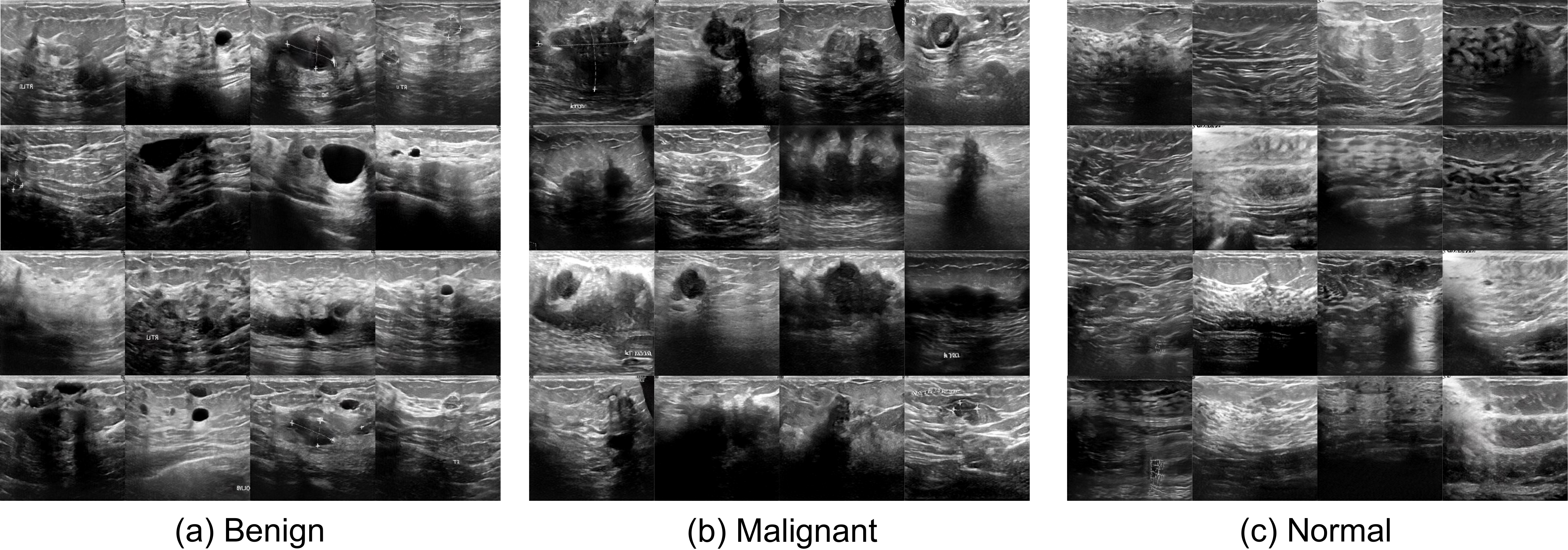}
	  \caption{Images synthesized by SGA with the TL technique. (a) benign, (b) malignant, and (c) normal.}
	  \label{fig3}
\end{figure*}

\begin{figure*}
	\centering
		\includegraphics[width=0.86\textwidth]{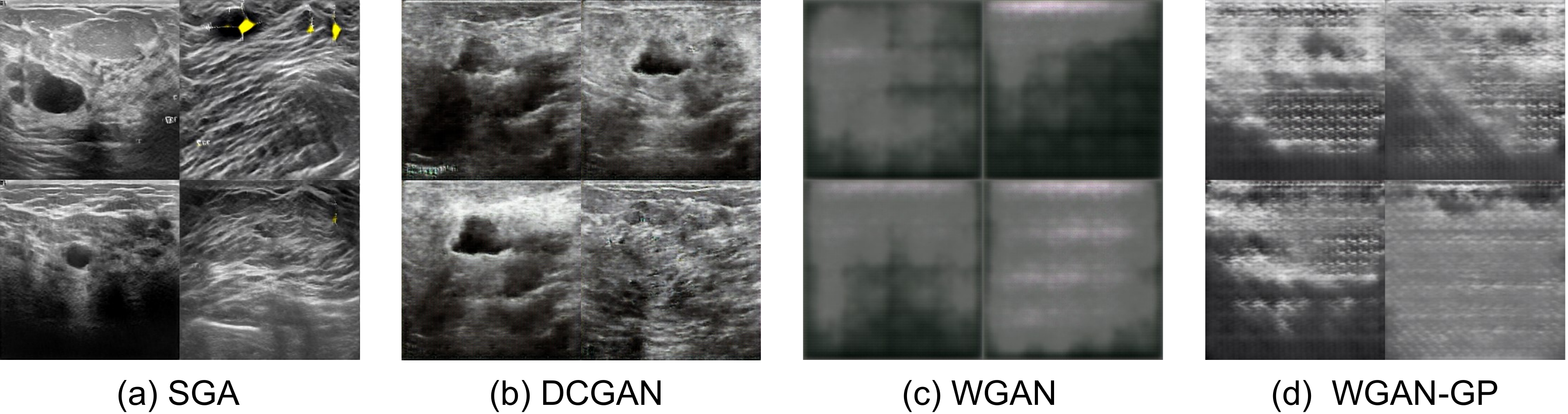}
	  \caption{Normal images synthesized by different GAN models. Models are trained from scratch. (a) SGA, (b) DCGAN, (c) WGAN, and (d) WGAN-GP.}
	  \label{fig4}
\end{figure*}

\begin{figure*}
	\centering
	  \includegraphics[width=0.9\textwidth]{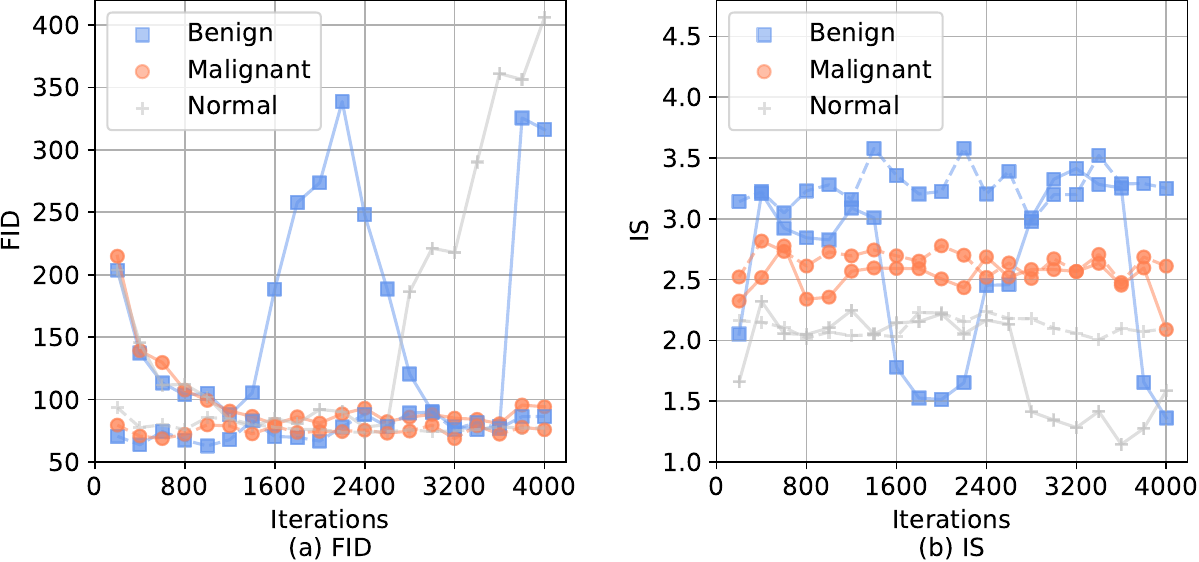}
	\caption{FID and IS recording during training. The solid lines stand TL from FFHQ, while dotted lines illustrate training from scratch. (a) FID, and (b) IS.}
	\label{fig5}
\end{figure*}

Several medical US images synthesized by SGA are shown in \Cref{fig3} and \Cref{fig4}a, for using the TL technique and training from scratch, respectively. We also show images synthesized by DCGAN, WGAN, and WGAN-GP in \Cref{fig4}b, \Cref{fig4}c, and \Cref{fig4}d for comparison. The synthesized images from SGA exhibit noticeably higher quality compared to those from other GAN models. The SGA effectively mimics the medical annotations (white in figures) in the BUSI dataset, while the commonly used DCGAN, WGAN, and WGAN-GP cannot even handle the task well at this resolution. Besides, the quality of the synthesized images significantly improves with the introduction of the TL technique. On a qualitative note, \Cref{fig3}a, \Cref{fig3}b, and \Cref{fig3}c do not exhibit the evident yellow flaws that are present in \Cref{fig4}a. Quantitatively, better FID and IS are observed, as shown in \Cref{fig5}a and \Cref{fig5}b, respectively. From the observation, we can find that the TL technique not only enhances performance but also improves stability. As the TL technique can improve the performance of SGA essentially, we set the TL technique as the default when developing SGA-CNN pairs. It is worth noting that the TL source dataset used here is FFHQ. For the comparison of FID and IS across different TL experimental groups, see the corresponding ablation study in Section \hyperref[4.3]{4.3}. 

\begin{figure}
	\centering
		\includegraphics[width=0.38\textwidth]{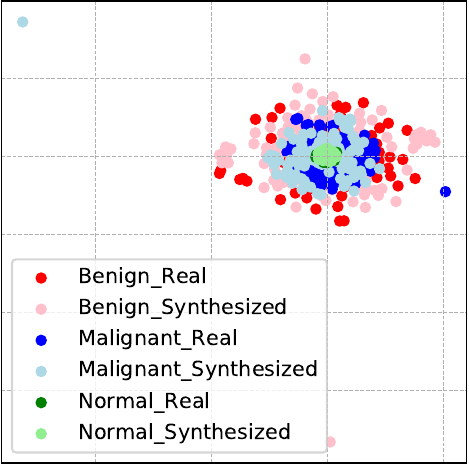}
	  \caption{Data contribution visualization using t-SNE across real and synthesized images.}
	  \label{fig6}
\end{figure}

\begin{figure*}
	\centering
	  \includegraphics[width=\textwidth]{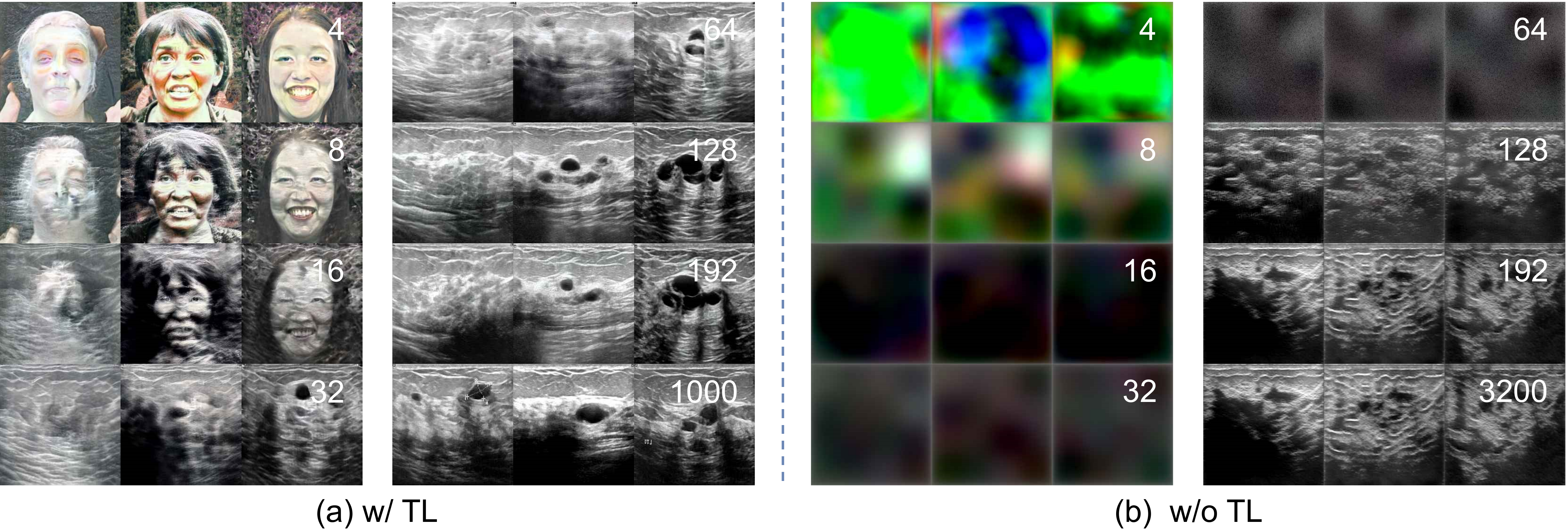}
	\caption{Images synthesized by SGA at different stages during training. Only for illustration. Numbers represent the number of iterations trained. (a): with TL, transfer from FFHQ, and (b): without TL, training from scratch.}
	\label{fig7}
\end{figure*}

To prove the effectiveness of the proposed image synthesis method, we employ t-SNE to visualize the data distribution across real and synthesized images in \Cref{fig6}. The visualization is performed using SGA-VGG without decay as it outperforms other combinations. Features are extracted prior to the classification head, and each category comprises a hundred randomly sampled images. The results demonstrate that the distributions of both real and synthesized images are closely aligned, and a nearly overlapping distribution attests to the effectiveness of the proposed synthesis method. Notably, several outliers are observed in both synthesized benign and malignant categories. This can be caused by either CNN prediction or SGA synthesize deviation. Nevertheless, the number of such outliers is limited and thus does not influence the overall results. In \Cref{fig7}, we illustrate how the TL technique aids the model training and the images synthesized during the process. When the TL technique is employed, as evident in \Cref{fig7}a, the $G$ of SGA inherits the weights learned from the FFHQ dataset. With pre-learned weights, the $G$ can learn the distribution of the BUSI dataset quickly. By the 32nd iteration, the $G$ can already synthesize the BUSI-like images. The lowest FID is reached at the 1000th iteration. However, in the case of lacking the TL technique, the weights are initialized randomly at the beginning of the training, as illustrated in \Cref{fig7}b. The $G$ starts to learn some representations at around 64 to 128 iterations, and the lowest FID is reached at the 3200th iteration. This indicates a substantially longer training time consumption compared to scenarios utilizing the TL technique. Worse still, even with 3200 iterations, the $G$ trained from scratch exhibits severe mode collapse. In other words, its diversity is significantly lower compared to that achieved with TL from FFHQ.

\subsection{Semi-supervised classification}
\label{4.2}

\begin{table}
\caption{Classification accuracy across SGA-CNN pairs. $macc$ shows the maximum accuracy across all $ \gamma $. $ \uparrow $ means the higher, the better, and $ \downarrow $ inverse. Bold numbers show the best results.}
\setlength{\tabcolsep}{2.4pt}
\resizebox{\linewidth}{!}{
\begin{tabular*}{572pt}{ ccccccccc }
\toprule
    WD         & Standard                  & SGA-VGG           & SGA-Shuffle & SGA-ResNeXt & SGA-Res & SGA-Mobile    & SGA-Inception & SGA-Dense \\
\midrule
    \checkmark & $ acc_{\gamma} \uparrow $ & 97.2\%            & 91.6\%      & 84.5\%      & 85.4\%  & 94.9\%        & 81.7\%        & 90.0\%  \\
    \checkmark & $ macc \uparrow $         & \textbf{97.3\%}   & 91.6\%      & 84.5\%      & 85.4\%  & 94.9\%        & 82.7\%        & 90.4\%  \\
    \checkmark & $ t_{\gamma} \downarrow $ & 1480.8s           & 1458.2s     & 1467.5s     & 1457.6s & 1460.3s       & 1385.6s       & 1263.7s \\
    \checkmark & $ \gamma $                & 7                 & 7           & 7           & 7       & 7             & 4             & 5       \\
    \checkmark & $ TEI_{\gamma} \uparrow $ & 2.18              & 2.75        & 2.55        & 2.14    & 2.86          & 2.28          & \textbf{3.06} \\
    $ \times $ & $ acc_{\gamma} \uparrow $ & 97.8\%            & 95.0\%      & 81.2\%      & 86.6\%  & 94.9\%        & 82.4\%        & 88.8\%  \\
    $ \times $ & $ macc \uparrow $         & \textbf{97.9\%}   &95.0\%       & 81.2\%      & 86.6\%  & 95.0\%        & 82.4\%        & 89.1\%  \\
    $ \times $ & $ t_{\gamma} \downarrow $ & 1478.4s           & 1456.5s     & 1251.9s     & 1455.5s & 1460.3s       & 1175.0s       & 1260.1s \\
    $ \times $ & $ \gamma $                & 7                 & 7           & 5           & 7       & 7             & 3             & 5       \\
    $ \times $ & $ TEI_{\gamma} \uparrow $ & 2.18              & 2.96        & 1.69        & 2.14    & \textbf{3.04} & 2.46         & 2.06    \\
\bottomrule 
\end{tabular*}
}
\label{tab1}
\end{table}

\begin{table}
    \centering
    \caption{Performance comparison on the BUSI dataset between GSDA and state-of-the-art methods. $^{*}$ shows binary classification. $^{**}$ stands including additional training data. SSL presents whether the methods belong to semi-supervised or not.}
    \setlength{\tabcolsep}{3.2pt}
    \resizebox{\linewidth}{!}{
    \begin{tabular*}{518pt}{ ccccc }
    \toprule
        Ref.                             & Year & SSL        & Methods                                                              & $ macc \uparrow $ \\
    \midrule 
        \citep{erouglu2021convolutional} & 2021 & $ \times $ & Multi-CNN Hybrid Structure                                           & 95.6\% \\
        \citep{das2021exploring}         & 2021 & $ \times $ & ResNet                                                               & 88.9\% \\
        \citep{khanna2021improving}      & 2021 & $ \times $ & ResNet + Binary Gray Wolf Optimization + Support Vector Machine      & 84.9\% \\
        \citep{joshi2022an}              & 2022 & $ \times $ & YOLO                                                                 & 95.3\% \\
        \citep{balaha2022hybrid}         & 2022 & $ \times $ & CNN + Genetic Algorithm $^{**}$                                      & 92.8\% \\  
        \citep{sahu2023high}             & 2023 & $ \times $ & ShuffleNet-ResNet $^{*}$                                             & 95.1\% \\
        \citep{wang2023information}      & 2023 & $ \times $ & Interpretable Multitask Information Bottleneck Network $^{*}$        & 93.0\% \\
        \citep{lei2023core}              & 2023 & $ \times $ & Consistent Ordinal Representations                                   & 82.2\% \\
        \citep{xu2023regional}           & 2023 & $ \times $ & Multi-Task Learning + Attention $^{*}$                               & 91.0\% \\
        \citep{gheflati2021vision}       & 2021 & $ \times $ & Vision Transformer                                                   & 74.0\% \\
        \citep{moon2020computer}         & 2020 & $ \times $ & CNN Ensemble Learning $^{*}$                                         & 90.8\% \\
        \citep{sadad2020identification}  & 2020 & $ \times $ & Hybrid Feature Set + Ensemble Classifier $^{*}$                      & 96.6\% \\
        \citep{mishra2021breast}         & 2021 & $ \times $ & Machine Learning-Radiomics $^{*}$                                    & 97.4\% \\
        \citep{byra2021breast}           & 2021 & $ \times $ & Deep Representations Scaling $^{*}$                                  & 92.3\% \\
        \citep{al2019deep}               & 2019 & \checkmark & CNN + DAGAN                                                          & 94.0\% \\
        \citep{xie2021dk}                & 2021 & \checkmark & ResNet + DK-Guided Data Augmentation                                 & 81.1\% \\
        \citep{song2022deep}             & 2022 & \checkmark & ResNet + Convolutional Autoencoder $^{*}$                            & 88.2\% \\
        \citep{wang2022semi}             & 2022 & \checkmark & Consistency Training + Vision Transformer + Adaptive Token Sampler   & 95.3\% \\
        Ours                             & -    & \checkmark & GSDA                                                                 & \textbf{97.9\%} \\
    \bottomrule 
    \end{tabular*}
    \label{tab2}}
\end{table}

The classification accuracy of different SGA-CNN pairs is shown in \Cref{tab1}. We find that the SGA-VGG pair without WD achieves the highest accuracy at 97.9\%. For the SGA-VGG pair with WD, we obtain an accuracy of 97.3\%. While the SGA-VGG pairs achieve the highest accuracy, several pairs demonstrate a greater improvement in accuracy with limited time consumption, reaching higher TEI. For instance, the SGA-Dense pair with WD reaches a TEI of 3.06, and the SGA-Mobile pair without WD obtains a TEI of 3.04. The intensive experiment results indicate that our method is universally suitable for all CNN models without any selection bias. We use the SGA-VGG pair without WD when comparing the performance with existing methods. In \Cref{tab2}, we compare our GSDA with the state-of-the-art methods using the same dataset. We divide the existing methods into two categories, depending on whether the method is semi-supervised or not. From the table, we can find that the proposed GSDA reaches the highest accuracy and overperforms existing methods, even for comparison with binary classification and training with extra data. This demonstrates the effectiveness of GSDA, establishing it as a new state-of-the-art milestone.

\begin{figure}
	\centering
		\includegraphics[width=0.9\textwidth]{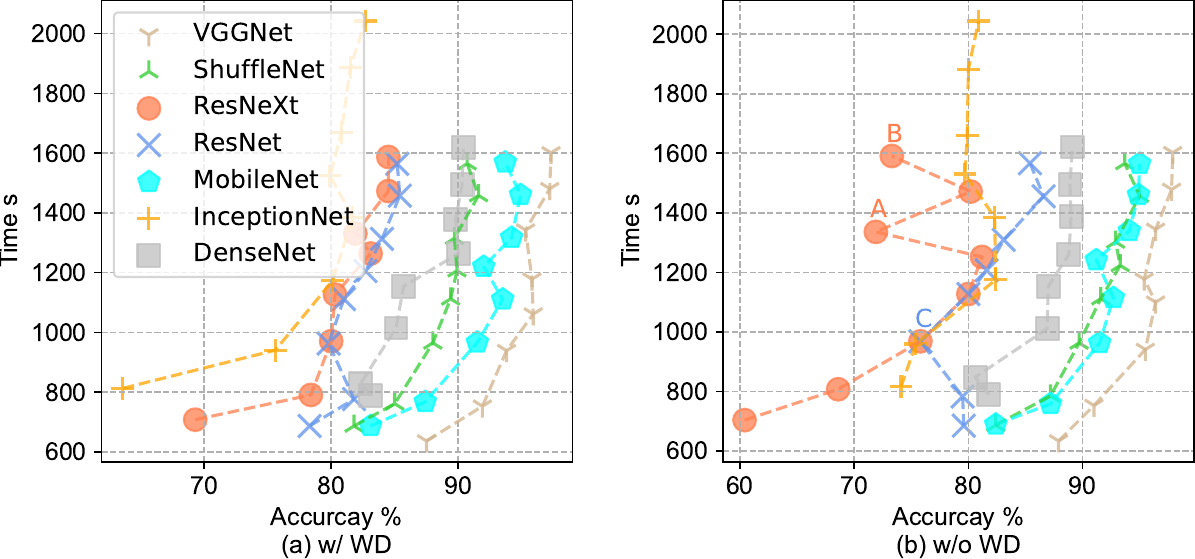}
	  \caption{Accuracy-time curve across different SGA-CNN pairs. The closer to the lower right corner, the better the overall performance. For each pair, the order of scatters corresponds to the increase of $ \gamma $. Two subplots share the legend. (a) with WD, and (b) without WD.}
	  \label{fig8}
\end{figure}

To provide guidance for practical applications, we plot the comparison of the accuracy-time curve across different SGA-CNN pairs in \Cref{fig8}a and \Cref{fig8}b, with and without the TL technique, respectively. It is worth noting that though the maximum value of $ \gamma $ varies among different SGA-CNN pairs, we conduct additional experiments and illustrate the results based on the maximum $ \gamma $ across all SGA-CNN pairs, which is eight as illustrated in \Cref{tab1}. In other words, each SGA-CNN pair has eight groups of experiments for both WD settings, respectively. This standardizes the results, making them more amenable to comparison. The results reveal that the SGA-VGG pair achieves the highest accuracy with the least time consumption, irrespective of the presence of WD. This illustrates that the SGA-VGG pair should be priority considered when deploying the classification task in practice. It is worth noting that the training of several pairs can become unstable without the WD, indicated in points A, B, and C in \Cref{fig8}b. This suggests that WD contributes to stabilizing the training to a certain degree.
 
\subsection{Ablation studies}
\label{4.3}

\begin{table}
\centering
\footnotesize
\caption{Classification accuracy across different CNN models.}
\setlength{\tabcolsep}{3.2pt}
\resizebox{\linewidth}{!}{
\begin{tabular*}{420pt}{ ccccccccc }
\toprule
    WD          & Standard                    & VGGNet           & ShuffleNet  & ResNeXt  & ResNet   & MobileNet  & InceptionNet  & DenseNet \\
\midrule
    \checkmark  & $ acc_{\gamma=0}\uparrow $  & \textbf{81.7\%}  & 72.2\%      & 66.5\%   & 70.3\%   & 74.7\%     & 65.8\%        & 69.0\%   \\
    \checkmark  & $ t_{\gamma=0}\downarrow $  & \textbf{262.7s}  & 290.8s      & 300.6s   & 291.0s   & 291.8s     & 327.4s        & 305.8s   \\
    $ \times $  & $ acc_{\gamma=0}\uparrow $  & \textbf{82.3\%}  & 74.1\%      & 69.6\%   & 71.5\%   & 73.4\%     & 65.8\%        & 74.7\%   \\
    $ \times $  & $ t_{\gamma=0}\downarrow $  & \textbf{260.5s}  & 290.0s      & 300.3s   & 291.8s   & 293.2s     & 327.3s        & 306.4s   \\
\bottomrule
\end{tabular*}
}
\label{tab3}
\end{table}

The structure ablation study is implemented by comparing the classification performance in the case of whether the SGA is implemented or not. The performance of the CNN models without SGA is detailed in \Cref{tab3}. A comparison between \Cref{tab1} and \Cref{tab3} reveals a significant drop in performance without SGA. For instance, the SGA-VGG pair shows a 15.6\% decrease in accuracy regardless of whether the WD is implemented. Given the limited size of the dataset, these results align with our expectations and underscore the effectiveness of SGA. Regarding the SGA-Inception pairs, a tremendous accuracy drop of 16.9\% and 16.6\% are observed in scenarios with and without WD, respectively.

\begin{table*}
    \centering
    \caption{Comparison of FID and IS across different TL experimental groups. In each group, the listed FID and IS are the optimal results calculated in the corresponding number of iterations.}
    \resizebox{0.8\linewidth}{!}{
    \begin{tabular*}{388pt}{ ccccccc }
    \toprule
        Group     & Subset     & TL        & FID $ \downarrow $  & Iterations  & IS $ \uparrow $  & Iterations \\
    \midrule
        1         & Benign     & CelebA    &  69.24              &  200        & \textbf{3.58}    & 2200 \\
        2         & Benign     & LSUN DOG  &  102.95             &  600        & 2.92             & 4000 \\
        3         & Benign     & FFHQ      &  \textbf{62.92}     &  1000       & \textbf{3.58}    & 2200 \\
        4         & Malignant  & CelebA    &  72.55              &  400        & \textbf{2.84}    & 2000 \\
        5         & Malignant  & LSUN DOG  &  89.68              &  600        & 2.25             & 600  \\
        6         & Malignant  & FFHQ      &  \textbf{68.78}     &  3200       & 2.82             & 400  \\
        7         & Normal     & CelebA    &  79.69              &  600        & 2.19             & 1200 \\
        8         & Normal     & LSUN DOG  &  91.63              &  200        & 2.01             & 1400 \\
        9         & Normal     & FFHQ      &  \textbf{73.92}     &  2200       & \textbf{2.24}    & 2400 \\
    \bottomrule
    \end{tabular*}
    }
    \label{tab4}
\end{table*}

The dataset ablation study is conducted by comparing FID and IS across various TL experimental groups. From \Cref{tab4}, it is found that TL from FFHQ performs best compared with TL from CelebA and TL from LSUN DOG, getting an FID of 62.92, 68.78, and 73.92 for three subsets respectively. However, for the malignant subset, TL from FFHQ obtains a lower IS compared with TL from CelebA. This conflicting outcome highlights some limitations of IS. It is worth noting that despite its higher diversity, LSUN DOG performs the worst in our experiments. This observation contrasts with the conclusion that the success of the TL technique likely hinges more on dataset diversity than on the similarity between subjects \citep{karras2020training}. We speculate that this conclusion might be influenced by the close relationship between dogs and cats.


\section{Conclusions}
\label{5}

We introduced the GSDA, a novel method aimed at enhancing the classification accuracy of medical US images under small data limits. Experimental results on the BUSI dataset underscore the effectiveness and robustness of GSDA in image classification. Given its commendable performance, GSDA has the promising potential to serve as a supplementary diagnostic instrument. However, it is imperative to acknowledge certain limitations. The SGA is trained independently on distinct subsets to mitigate mutual interference for performance consideration. However, when there are numerous subsets, this approach may become impractical due to computational resource constraints. In such scenarios, the SGA can be conditionally trained by feeding class labels alongside the images. Using the trained SGA, images from various subsets can then be synthesized. Furthermore, a potential challenge of GSDA is the complexity introduced by separately training the SGA and CNN. To mitigate this, the two stages can be trained synchronously, leveraging the $D$ of SGA for classification. While this training method allows for synchronous training of SGA and CNN, it poses challenges when comparing performances across diverse CNN models. Integrating CNN into SGA demands significant computational resources, given that the computational cost of SGA surpasses that of CNN by multiple orders of magnitude. The avenues for future research can be categorized into three primary domains. Firstly, while the GSDA is designed for 2D medical image classification, there is potential to extend the method to image segmentation and 3D imaging. Secondly, the GSDA presently sets the size of the extended dataset through comprehensive experimentation. Exploring more efficient methods to determine this size could curtail computational costs. Lastly, in light of the rapid advancements in the vision transformer \citep{dosovitskiy2020image}, integrating CNN and the vision transformer appears promising. Such integrations can potentially enhance model performance by effectively capturing both local and global features, as discussed in \citep{liu2023recent}.





\bibliographystyle{unsrt}
\bibliography{reference.bib}

\biboptions{sort&compress}







\end{sloppypar}
\end{document}